\documentclass[prl,amsmath,amssymb,showpacs,floatfix,footinbib,twocolumn,superscriptaddress]{revtex4-1}

\usepackage[english]{babel}
\usepackage[utf8x]{inputenc}
\usepackage{graphicx}
\usepackage{amsmath}
\usepackage{amssymb}
\usepackage{xspace}

\let\AAt\AA
\renewcommand{\AA}{\ifmmode{\mathrm{\AAt}}\else{\AAt}\fi}
\newcommand{\var}[1]{\ensuremath{#1}\xspace}
\newcommand{\unt}[1]{\ensuremath{\mathrm{#1}}\xspace}

\newcommand{\chem}[1]{\ensuremath{\rm #1}\xspace}
\newcommand{\mB}{{\rm \mu_B}}
\newcommand{\Jr}{J\!\rho_0}
\newcommand{\hBN}{{\it h}\text{--}BN}

\newcommand{\ket}[1]{\left\vert #1 \right\rangle}
\newcommand{\abs}[1]{\left|#1\right|}

\preprint{CoH/BN/Rh(111) v1.0}

\begin{document}

	\title{Quantum Engineering of Spin and Anisotropy in Magnetic Molecular Junctions}
	
	\author{Peter Jacobson$^\star$}
	\email{p.jacobson@fkf.mpg.de}
	\affiliation{Max Planck Institute for Solid State Research, 
Heisenbergstrasse 1, 70569 Stuttgart, Germany}
	\author{Tobias Herden$^\star$}
	\affiliation{Max Planck Institute for Solid State Research, Heisenbergstrasse 1, 70569 Stuttgart, Germany}
	\author{Matthias Muenks}
	\affiliation{Max Planck Institute for Solid State Research, Heisenbergstrasse 1, 70569 Stuttgart, Germany}
	\author{Gennadii Laskin}
	\affiliation{Max Planck Institute for Solid State Research, Heisenbergstrasse 1, 70569 Stuttgart, Germany}
	\author{Oleg Brovko}
	\affiliation{Max Planck Institute of Microstructure Physics, Weinberg 2, 06120 Halle, Germany}
	\author{Valeri Stepanyuk}
	\affiliation{Max Planck Institute of Microstructure Physics, Weinberg 2, 06120 Halle, Germany}
	\author{Markus Ternes}
	\email{m.ternes@fkf.mpg.de}
	\affiliation{Max Planck Institute for Solid State Research, Heisenbergstrasse 1, 70569 Stuttgart, Germany}
	\author{Klaus Kern}
	\affiliation{Max Planck Institute for Solid State Research, Heisenbergstrasse 1, 70569 Stuttgart, Germany}
	\affiliation{Institute de Physique de la Matière Condensée, École Polytechnique Fédérale de Lausanne, 1015 Lausanne, Switzerland}

 \begin{abstract}
 	Single molecule magnets and single spin centers can be individually 
addressed when coupled to contacts forming an electrical junction. In order to 
control and engineer the magnetism of quantum devices, it is necessary to 
quantify how the structural and chemical environment of the junction affects the 
spin 
center \cite{Bogani2008,Gatteschi2003,Rau2014,Miyamachi2013,Wegner2009,
Heinrich2013}. Metrics such as coordination number or symmetry provide a simple 
method to quantify the local environment, but neglect the many-body interactions 
of an impurity spin when coupled to contacts \cite{Oberg2014}. Here, we utilize 
a highly corrugated hexagonal boron nitride (\chem{\hBN}) 
monolayer~\cite{Laskowski2007,Herden2014} to mediate the coupling between a 
cobalt spin in \chem{CoH_x} (\chem{x=1,2}) complexes and the metal contact. 
While the hydrogen atoms control the total effective spin, the corrugation is 
found to smoothly tune the Kondo exchange interaction between the spin and the 
underlying metal. Using scanning tunneling microscopy and spectroscopy together 
with numerical simulations, we quantitatively demonstrate how the Kondo exchange 
interaction mimics chemical tailoring and changes the magnetic anisotropy.\\ \\
$^\star$P.\ J. and T.\ H. contributed equally.
 \end{abstract}

	\maketitle

	Magnetic anisotropy defines the stability of a spin in a preferred 
direction \cite{Gatteschi2003}. For adatoms on surfaces, the low coordination 
number and changes in hybridization can lead to dramatic enhancement of magnetic 
anisotropy \cite{Rau2014,Gambardella2003}. Surface adsorption site and the 
presence of hydrogen has been shown to alter the magnetic anisotropy of adatoms 
on bare and graphene covered 
\chem{Pt(111)} \cite{Khajetoorians2013,Dubout2015,Donati2013}. Furthermore, the 
exchange interaction and strain has been invoked for $3d$ adatoms on 
\chem{Cu_2N} islands where the adatom position on the island affects the 
observed magnetic anisotropy \cite{Oberg2014,Bryant2013}. Studies on single 
molecule magnets (SMMs) containing $3d$ or $4f$ spin centers have revealed that 
chemical changes to the ligands surrounding the spin affect the magnetic 
anisotropy \cite{Jurca2011}. However, the most important factor for maintaining 
magnetic anisotropy in SMMs is a low coordination number and a high axial 
symmetry \cite{Miyamachi2013,Zadrozny2013,Ungur2014}.

	Magnetic anisotropy is not guaranteed in SMMs or single spin centers 
upon coupling to contacts. The spin interacts with the electron bath through 
the exchange interaction leading to a finite state lifetime and the decay of 
quantum coherence \cite{Cohen-Tannoudji2008,Delgado2014c}. Additionally, the 
scattering of the spin with the electron bath results in an energy 
renormalization of the spin's eigenstate energy levels, similar to the case of a 
damped harmonic oscillator \cite{Cohen-Tannoudji2008}. In practice, this leads 
to a net reduction of the magnetic anisotropy, pushing the system closer to a 
Kondo state. At the heart of the Kondo effect are spin--flip scattering 
processes between localized states at the impurity spin and delocalized states 
in the bulk conduction band, resulting in the formation of a correlated quantum 
state \cite{Hewson1993}. The Kondo regime is reached when the magnetic moment 
of the impurity spin is screened by the electron bath, with the exchange 
interaction defining the relevant energy scale, the Kondo temperature 
(\var{T_K}). High spin systems with a total spin $\var{S} > 1/2$ have the 
potential for both magnetic anisotropy and the Kondo 
effect \cite{Misiorny2012,Zitko2008}. Thus, the Kondo exchange interaction with 
the electron bath can force the impurity spin into a competing Kondo state, 
where antiferromagnetic coupling with the reservoir reduces or even quenches the 
magnetic moment. The outcome of this competition can be determined in local 
transport measurements, but few quantitative measures of this competition exist.

	Here, we study \chem{CoH_x} complexes coupled to a spatially varying 
template, the \chem{\hBN/Rh(111)} moiré, to observe and model how the 
environment influences magnetic anisotropy. The \chem{\hBN} monolayer, a wide 
band gap two dimensional material, decouples and mediates the interactions 
between \chem{CoH_x} and the underlying \chem{Rh} metal while lattice mismatch 
leads to a spatial corrugation resulting in an enlarged unit cell with 3.2 nm 
periodicity corresponding to 13 \chem{BN} units on top of 12 \chem{Rh} 
atoms \cite{Laskowski2007}. The local adsorption configuration of \chem{CoH_x} 
on the \chem{\hBN} is conserved across the moiré unit cell, with the large 
number of inequivalent adsorption sites allowing us to explore how hybridization 
affects magnetic anisotropy. To complement our experimental observations, we 
model transport through the \chem{CoH_x} complexes using Hamiltonians that 
incorporate magnetic anisotropy as well as coupling to the environment. This is 
accomplished by parameterizing the environment through use of a dimensionless 
coupling constant $-\Jr$, describing the strength of the Kondo exchange 
interaction, $\var{J}$, between the localized spin and the electron density 
$\rho_0$ of the substrate near the Fermi level (see SI).
	
	\begin{figure}
		\center{\includegraphics{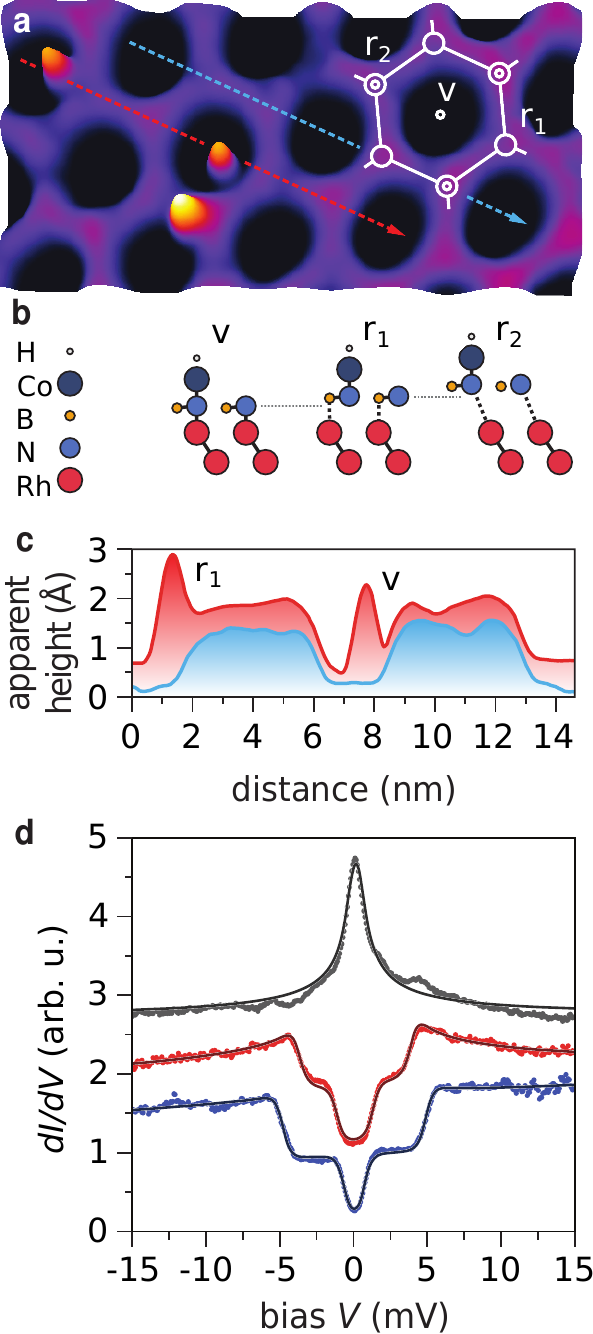}}
		\caption{\textbf{CoH$_x$ adsorbed on a
 \textit{h}--BN/Rh(111) surface.} (a) Constant current STM topography with 
three CoH$_x$ complexes (protrusions) adsorbed on different sites 
($15\times7~\unt{nm^2}$ image size, $\var{V} = -100~\unt{mV}$, $\var{I} =  
20~\unt{pA}$, $\var{T} = 1.4~\unt{K}$). High symmetry points of the moiré are  
marked by the white overlay. (b) Sketch of the atom positions for the adsorption 
of CoH. The $h$-BN registry with Rh(111) shifts across the 
moiré unit cell with three high symmetry sites: at the valley site (v) 
the Rh is directly underneath the N, whereas for the two unequal 
rim sites (r$_1$ and r$_2$) changes in the registry and distance to 
the surface are observed. (c) Line profiles along the dashed lines indicated 
in (a) show two CoH$_x$ systems with adsorption sites r$_1$ and 
v (red line) and a $h$-BN reference cut (blue line, offset by 
$0.5~\AA$). (d) Differential conductance \var{dI/dV} curves versus bias 
voltage of three different CoH$_x$ systems (stabilization setpoint: 
$\var{I} =  500~\unt{pA}$, $\var{V} = -15~\unt{mV}$, $\var{T} = 1.4~\unt{K}$, 
curves vertically offset for clarity). The upper curve (grey) shows a 
spin--$1/2$ Kondo resonance centered at zero bias. The two lower curves (red 
and blue) show step-like conductance increases symmetric around zero bias 
indicating a spin--1 system. Solid black lines are least-square fits using a 
perturbative transport model.}
		\label{fig:01:stm}
	\end{figure}

	Figure~\ref{fig:01:stm}a shows a representative scanning tunneling 
microscopy (STM) topograph of the \chem{\hBN/Rh(111)} moiré with isolated 
\chem{CoH_x} (\chem{x=1,2}) complexes, line profiles across the \chem{\hBN} 
indicate \chem{CoH_x} can adsorb at multiple positions within the moiré 
(Figure~\ref{fig:01:stm}b) \cite{Natterer2012}. On these \chem{CoH_x} complexes 
we measure the differential conductance, \var{dI/dV}, against the applied bias 
voltage \var{V} between tip and sample at low-temperature ($\var{T} = 
1.4~\unt{K}$) and zero magnetic field ($\var{B} = 0~\unt{T}$, details see 
Methods). The spectra can be divided into two broad classes: a sharp peak 
centered at zero bias or two symmetric steps of increasing conductance at 
well-defined threshold energies (Figure~\ref{fig:01:stm}c). The peak at zero 
bias is consistent with a spin--$1/2$ Kondo resonance while the steps 
correspond to the onset of inelastic excitations from the magnetic ground state 
to excited states. The observation of two steps hints at a spin--1 system with 
zero field splitting. The two lower spectra (Figure~\ref{fig:01:stm}c; red, blue 
curves) are measured on CoH at different parts of the moiré and share the 
same characteristics but the step positions vary.

	\begin{figure*}
		\center{\includegraphics{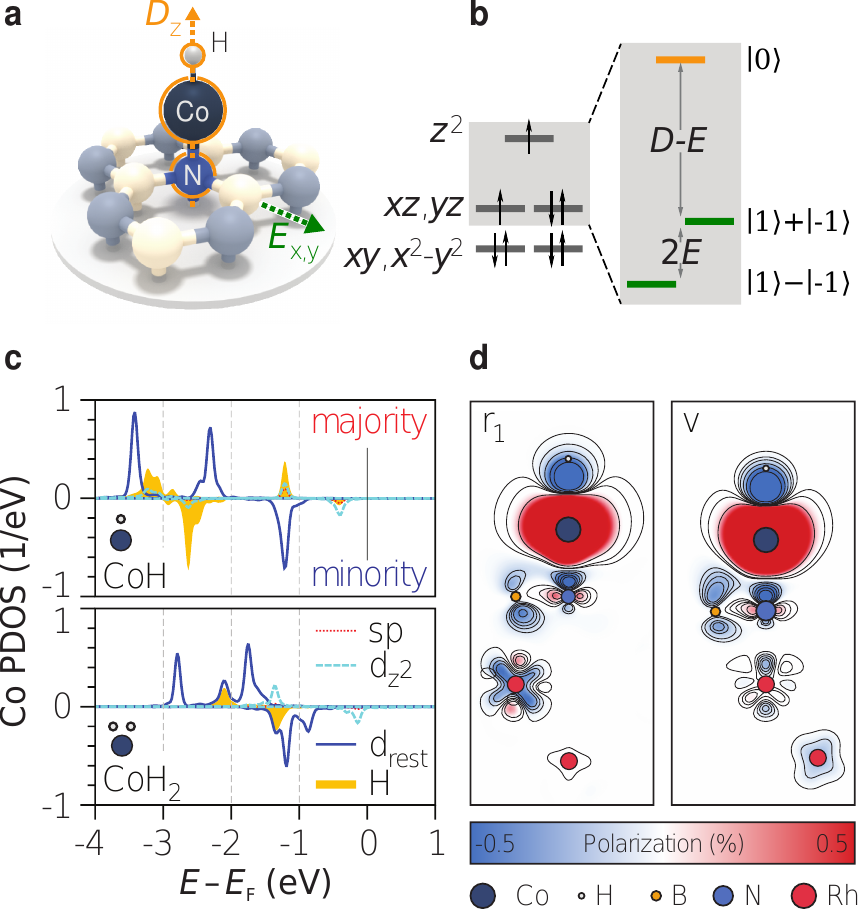}}
		\caption{\textbf{CoH and CoH$_2$ density of 
states.} (a) Ball and stick model of the adsorption of CoH on 
$h$-BN. The linear adsorption geometry of CoH on the N atom 
is emphasized and marks the main (axial) magnetic anisotropy (\var{D}) along the 
$z$-axis. Additional transverse anisotropy ($E$) in the $x-y$ plane further 
breaks the symmetry. (b) Schematic linear crystal field splitting diagram for 
the $3d^8$ shell of Co highlighting the origin of the axial (\var{D}) 
and transverse (\var{E}) magnetic anisotropy. The magnetic ground state is an 
antisymmetric superposition of $\var{m_z} = \ket{+1}$ and $\ket{-1}$ states 
(\var{m_z} is the magnetic moment in units of the reduced Planck constant 
$\hbar$ in \var{z}-direction), the first excited state is the symmetric 
superposition, and the second excited state is $\var{m_z}=\ket{0}$. (c) Plots of 
the majority and minority spin projected density of states (PDOS) for CoH 
and CoH$_2$. The difference in majority and minority spin spectral weights 
indicate that CoH has a total spin $\var{S}=1$ and CoH$_2$ has 
$\var{S} = 1/2$. (d) Plot of the asymmetry between majority and minority PDOS 
for CoH adsorbed on N at the r$_1$ (left) and v (right) high symmetry points.}
		\label{fig:02:theo}
	\end{figure*}

	We employ density functional theory (DFT) to correlate the magnetic 
properties of \chem{CoH_x} with the local adsorption configuration. Our 
calculations (see Methods) show that adsorption in the BN hexagon, \textit{i.e.} 
hollow site, is preferable for bare Co. The addition of hydrogen shifts 
the preferred adsorption site to N, with the hollow site adsorption 
energy consistently higher. For CoH complexes the preferred hydrogen 
position was found to be either exactly on top of Co or tilted towards 
the nearest \chem{B} atom (Figure~\ref{fig:02:theo}a). An important consequence 
of the N adsorption site is the linear crystal field acting on the cobalt 
(\textit{i.e.} \chem{N-Co-H}) removing the 5-fold degeneracy of the $d$-levels 
(Figure~\ref{fig:02:theo}b). 

	In Figure~\ref{fig:02:theo}c the spin--resolved, symmetry decomposed 
local density of states of CoH and CoH$_2$ adsorbed in the 
\chem{\hBN} valley is plotted. The atomic $d$-levels are split roughly 
$1.2~\unt{eV}$ by the intrinsic Stoner exchange giving a bare Co adatom a 
magnetic moment of 2.2 Bohr magnetons ($\mB$). Formation of CoH leads to 
hybridization of the H $sp$ orbitals and the Co orbitals, slightly 
reducing the magnetic moment to $2.0~\mB$, equivalent to a $3d^8$ configuration 
(Figure~\ref{fig:02:theo}b). The second hydrogen changes the picture 
significantly, with the $sp-d$ hybridization sufficient to bring the Co 
$d$-levels closer together, reducing the magnetic moment to $1.2~\mB$ resulting 
in a $3d^9$ configuration. Therefore, from our spectroscopic observations and 
DFT calculations we identify CoH as an effective spin--1 and CoH$_2$ 
as spin--$1/2$ system.

	Figure~\ref{fig:02:theo}d shows the spin density distribution for 
CoH in a N adsorption configuration at two high symmetry points of 
the moiré. The strong vertical bond between Co and N leads to an 
effective spin--polarization along this axis and can be expected to provide the 
system with out-of-plane magnetic anisotropy. Tilting of the hydrogen and the 
underlying lattice mismatch reduces the C$_{3v}$ symmetry and introduces 
small shifts in the $d_{xz}$, $d_{yz}$ levels producing a non-negligible 
in-plane component of the anisotropy. 

	To model the experimentally observed tunneling spectra and to determine 
the magnetic anisotropy we use a phenomenological spin Hamiltonian including the 
Zeeman energy and magnetic anisotropy:
	\begin{equation}
		\hat{H} = g \mB \; \vec{\boldsymbol{B}} \cdot \boldsymbol{\hat{S}} + D \hat{S}_z^2 + E (\hat{S}_x^2 - \hat{S}_y^2), \label{eq:ham}
	\end{equation}
	with \var{g} as the gyromagnetic factor, $\vec{\boldsymbol{B}}$ the 
magnetic field, $\boldsymbol{\hat{S}} = (\hat{S}_x, \hat{S}_y, \hat{S}_z)$ the 
total spin operator, and \var{D} and \var{E} as the axial and transverse 
magnetic 
anisotropy \cite{Oberg2014,Bryant2013,Otte2008,Hirjibehedin2007,Lorente2009}. 
Transport through the junction is calculated using a Kondo-like interaction 
$\hat{\sigma}\cdot\hat{S}$ between the tunneling electrons and the localized 
spin system, with $\hat{\sigma}$ as the standard Pauli matrices. We account for 
scattering up to $\mathrm{3^{rd}}$ order in the matrix elements by considering 
additional exchange processes between the localized spin and substrate electrons 
of the form~\cite{Zhang2013} (see SI):
	\begin{equation}
		\frac{1}{2}\Jr \; \hat{\sigma} \cdot \hat{S}.
	\end{equation}
	
	\begin{figure*}
		\center{\includegraphics{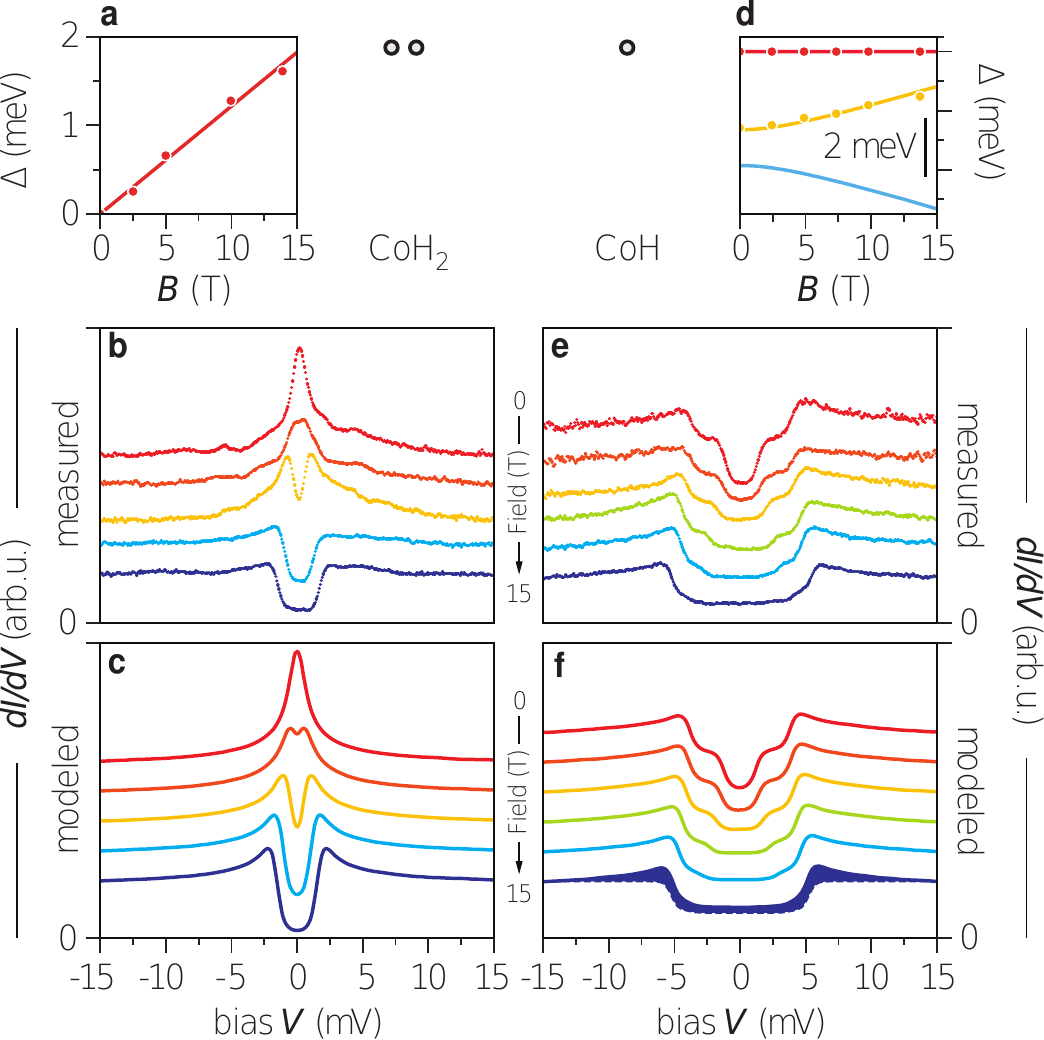}}
		\caption{\textbf{Magnetic field behavior of CoH$_2$ 
and CoH.} (a) Left: Zeeman splitting of the spin--$1/2$ states of a 
CoH$_2$ complex in magnetic field. Dots mark the energy differences as 
determined by least-square fits of the perturbation model to the experimental 
data in (b). The regression line corresponds to a gyromagnetic factor 
$\var{g}=2.0 \pm 0.1$. Right: Sketch of the CoH$_2$ complex adsorbed on a 
N site. (b) Evolution of the differential conductance of a CoH$_2$ 
complex in an external magnetic field normal to the surface ($\var{B} = 0, 2.5, 
5, 10$ and $14~\unt{T}$; $\var{T} = 1.4~\unt{K}$). (c) Simulated spectra using a 
$\mathrm{3^{rd}}$ order perturbation model and a constant coupling to the 
substrate of $-\Jr = 0.1$ and $\var{g} = 2.0$. (d) Left: Sketch of the spin--1 
CoH complex adsorbed on a N site. Right: State energy evolution in 
magnetic field along the out-of-plane anisotropy axis. Dots mark the 
experimentally determined step positions, full lines are the calculated 
eigenstate energies of the model Hamiltonian (see text) using magnetic 
anisotropy parameters of $\var{D} = -4.8~\unt{meV}$, $\var{E} = 0.6~\unt{meV}$, 
and $\var{g} = 2.2$. (e) Evolution of the differential conductance of a 
CoH system in an external magnetic field normal to the surface ($\var{B} 
= 0, 2.5, 5, 7.5, 10$ and $14~\unt{T}$; $\var{T} = 1.4~\unt{K}$). (f) Simulated 
spectra using the parameter from (d) and $-\Jr = 0.1$. The $14~\unt{T}$ spectrum 
is shown together with a $\mathrm{2^{nd}}$ order perturbation theory model, 
\textit{i.e.} $-\Jr = 0$ (dashed line), to highlight the necessity of 
$\mathrm{3^{rd}}$ order contributions. Curves in (b, c) and (e, f) are shifted 
vertically for better visibility.}
		\label{fig:03:field}
	\end{figure*}

	To confirm the magnetic origin of the spectroscopic features in 
CoH and CoH$_2$, we measure the differential conductance at magnetic 
fields up to $14~\unt{T}$ normal to the surface. Figure~\ref{fig:03:field}b 
shows experimental spectra taken over one CoH$_2$ complex and 
Figure~\ref{fig:03:field}c the model calculations for the Kondo resonance. 
Applying an external magnetic field introduces Zeeman splitting to the 
spin--$1/2$ system (Figure~\ref{fig:03:field}a). At low magnetic fields, 
$2.5~\unt{T}$, the peak broadens and the differential conductance of the 
resonance is reduced. Increasing the field to $5~\unt{T}$, a clear splitting of 
the Kondo resonance is observed. For the highest fields, the degeneracy of the 
spin--$1/2$ state is effectively lifted, resulting in a strong reduction of the 
Kondo resonance and the appearance of an inelastic excitation gap. We can 
reproduce the peak and its splitting by our perturbative model 
(Figure~\ref{fig:03:field}c) even though at high fields the peak-like 
conductance is weaker in the experimental data than expected from the model 
calculation. This indicates that the Kondo temperature of the system lies close 
to the base temperature of our experiment.
	
	Increasing the external magnetic field has two effects on the spin--1 
CoH; Zeeman splitting separates the steps and the ratio between inner and 
outer conductance step height decreases (Figure~\ref{fig:03:field}e). At zero 
field, the ground and first excited states are a superposition of $\var{m_z} = 
\ket{+1}$ and $\ket{-1}$ states, applying a magnetic field reduces the spin 
mixing and leads towards a $\ket{+1}$ ground and $\ket{-1}$ excited state. This 
accounts for the reduction of the inner step with increased magnetic field, as 
the transition between ground and first excited state becomes less probable 
because it would require a change in \var{m_z} of two. Reverting to a purely 2nd 
order simulation, large deviations are observed at both steps, evidence that 
coupling of the spin to the substrate conduction electron bath must be 
considered (Figure~\ref{fig:03:field}f, dashed line). The experimental data fits 
excellently when including 3rd order terms, \textit{i.e.} assuming a finite 
$-\Jr$, an out-of-plane anisotropy axis, and $\var{g} = 2.2 \pm 0.2$.

	\begin{figure*}
		\center{\includegraphics{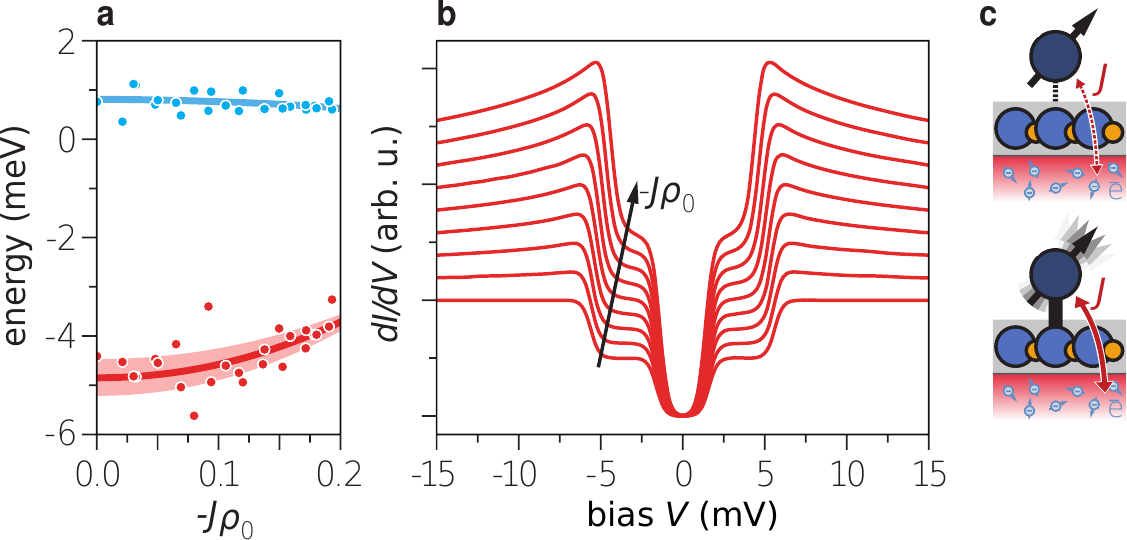}}
		\caption{\textbf{Influence of environmental coupling on 
\chem{\bf CoH} spectra.} (a) Experimentally determined \var{D} and \var{E} (red 
and blue dots) parameters plotted versus the coupling strengths $-\Jr$. Full 
lines show the expected renormalization of \var{D} and \var{E} due to virtual 
coherences calculated with a Bloch-Redfield approach taking exchange scattering 
with the dissipative substrate electron bath into account. Shaded region shows 
the experimental uncertainty. (b) Computed differential conductance for 
different coupling strengths between the localized spin and the electrons of the 
substrate ranging from $-\Jr = 0$ to $-\Jr = 0.2$. At stronger couplings 
($-\Jr>0$) an increase of the outer step's shoulder is expected concomitant with 
a reduction of the energy position of the outer step. This is equivalent to a 
reduced anisotropy energy \var{D}. (c) Schematic diagram showing the effect of 
exchange. When the exchange coupling, \var{J}, between the local spin and the 
conduction electron bath is weak, a large magnetic anisotropy, \var{D}, is 
observed (top). As exchange coupling to the substrate strengthens, the magnetic 
anisotropy is reduced driving the system closer to a Kondo state (bottom).}
		\label{fig:04:rhoj}
	\end{figure*}
	
	Evaluation of more than 30 CoH shows no sharp distribution of 
the anisotropy parameters $D$ and $E$. A transition of the main anisotropy axis 
into the surface plane occurs when $3\var{E} > \abs{\var{D}}$, therefore we 
have only considered complexes with a clear out-of-plane anisotropy determined 
by the criterion $\abs{\var{D}}/3\var{E} > 1.5$; a representative spectrum with 
in-plane anisotropy is shown in the SI. By considering the values of 
$-J\rho_0$ from our fits, we observe a correlation between the magnetic 
anisotropy and coupling with the substrate, $-J\rho_0$. The red branch in 
Figure~\ref{fig:04:rhoj}a shows that as the substrate coupling increases, the 
axial magnetic anisotropy decreases. These results are in line with predictions 
that increased coupling shifts energy levels. The solid red line shows the best 
fit to our data and follows the trend $\var{D} = \var{D}_0 
(1-\alpha(J\rho_0)^2)$, where $\alpha$ is a constant describing the bandwidth 
of the Kondo exchange interaction. The shaded red region accounts for the 
possible range of $\alpha$ by considering an effective bandwidth of $\omega_0 = 
0.4 - 1.2~\unt{eV}$ (see SI). For $0.1 < -J\rho_0 < 0.2$ the variation in 
magnetic anisotropy fits exceptionally well, but for small values of 
$-J\rho_0$, some spread in the axial anisotropy is observed. These 
fluctuations are not accounted for in our model and indicate that for small 
$-J\rho_0$ additional factors such as strain or defects may contribute to the 
magnetic anisotropy. While the axial anisotropy shows clear dispersion, the 
transverse anisotropy is essentially constant (Figure~\ref{fig:04:rhoj}a, blue).

Figure~\ref{fig:04:rhoj}b shows the influence of $-J\rho_0$ on the tunneling 
spectra calculated using a Bloch-Redfield approach to incorporate virtual 
correlations between the ground and excited states due to the coupling with the 
dissipative spin bath in the substrate assuming a flat density of states and an 
effective bandwidth of $\omega_0 = 1~\unt{eV}$ (see 
SI) \cite{Oberg2014,Cohen-Tannoudji2008,Delgado2014c}. As $-J\rho_0$ is 
increased, virtual correlations lead to renormalization and reduce the level 
splitting. This is observed experimentally as a reduction of the axial magnetic 
anisotropy. Furthermore, higher order scattering processes in the tunneling 
influence the conductance leading to an enhanced shoulder at the outer energy 
step that changes the contours of the spectrum (see SI). The symmetric peaks 
shift towards zero bias as $-J\rho_0$ increases indicating that correlations 
drive the anisotropic spin--1 system closer to the Kondo state. 
Figure~\ref{fig:04:rhoj}c schematically depicts the observed trend, when the 
spin is weakly coupled to the conduction electrons the magnetic anisotropy is 
stabilized. Increasing the exchange interaction introduces correlations between 
the excited spin states and the conduction electrons, leading to a net reduction 
in the magnetic anisotropy.

In conclusion, our results show that the Kondo exchange interaction modulates 
the magnetic anisotropy of single spin CoH complexes. The role of exchange was 
quantitatively determined by exploiting the corrugated \chem{\hBN} moiré 
structure. In conjunction with 3rd order perturbation theory simulations, we 
extracted the precise values of the spin coupling to the environment and its 
influence on the magnetic anisotropy. Kondo exchange must be considered an 
additional degree of freedom -- beyond local symmetry, coordination number, and 
spin state -- for spins connected to contacts. This parameter is non-local and 
therefore expected to be discernable at surfaces, in junctions, and perhaps in 
bulk SMM materials.\\
	
\section*{Acknowledgements}
	P.J. acknowledges support from the Alexander von Humboldt Foundation. 
T.H., M.M. and M.T. acknowledge support by the SFB 767. O.B. and V.S. 
acknowledge support by the SFB 762.

\section*{Author contributions}
	M.T. and K.K. conceived the experiments. T.H., P.J., M.M., and G.L. 
performed the STM measurements. P.J. and T.H. analyzed the data using a 
perturbation theory simulation package developed by M.T. O.B. and V.S. performed 
first principles density functional theory calculations. P.J., M.T., T.H., and 
O.B. drafted the manuscript; all authors discussed the results and contributed 
to the manuscript. 

\section*{Methods}
	
	The \chem{Rh(111)} surface was prepared by multiple cycles of argon ion 
sputtering and annealing to $1100~\unt{K}$. On the final annealing cycle 
borazine (\chem{B_3N_3H_6}) was introduced at a pressure of 
$1.2\times10^{-6}~\unt{mbar}$ for 2 minutes resulting in a monolayer \chem{\hBN} 
film. Cobalt was deposited onto a cold, $\sim20~\unt{K}$, \chem{\hBN} surface 
via an electron beam evaporator.
	
	Scanning tunneling experiments were performed on a home-built STM/AFM 
in ultra-high vacuum with a base temperature of $1.4~\unt{K}$ and magnetic 
fields up to $14~\unt{T}$. All spectroscopic (\var{dI/dV}) measurements 
presented were obtained with an external lock-in amplifier and a modulation 
voltage of $0.2~\unt{mV}$ applied to the bias voltage at a frequency of 
$799~\unt{Hz}$. The tunneling setpoint before the feedback loop was disabled was 
$\var{V} = -15~\unt{mV}$ and $\var{I} = 500~\unt{pA}$. For measurements on the 
same adatoms in different external magnetic fields the tip was retracted while 
the field was ramped and allowed to settle for maximum stability. 

First principles calculations have been carried out in the framework of the 
density functional theory (DFT) as implemented in the VASP 
code \cite{Kresse1993,Kresse1996}. We use the projector augmented-wave 
technique~\cite{Blochl1994} where the exchange and correlation were treated with 
the gradient-corrected PBE functional as formalized by Perdew, Burke and 
Ernzerhof \cite{Perdew1996}. Hubbard \var{U} and \var{J} values were taken from 
self-consistent calculations and fitting to experiments to be 
$\var{U}-\var{J}=3~\unt{eV}$ \cite{Steiner1992,Osterwalder2001,Wehling2010}. 
Full details are presented in the Supplementary Information.

	\bibliography{cobn}

\end{document}